\documentclass[aps,prl,floats,twocolumn,superscriptaddress,amssymb,amsmath,showpacs]{revtex4}
\usepackage{graphicx}
\usepackage{epstopdf}
\begin{document}
%\title{Thinning of superfluid films: critical effects immediately below the $\lambda$ point}
\title{Thinning of superfluid films below the critical  point}

\author{Roya Zandi}
\affiliation{Department of Physics and Astronomy, University of California, Riverside
900 University Avenue,
Riverside, CA 92521}

\author{Aviva Shackell}
\author{Joseph Rudnick}
\affiliation{Department of Physics and Astronomy, UCLA, Box 951547, Los Angeles,
CA 90095-1547}

\author{Mehran Kardar}
\affiliation{Department of Physics, Massachusetts Institute of
Technology, Cambridge, MA 02139, USA}

\author{Lincoln P. Chayes}
\affiliation{Department of Mathematics, UCLA, Box 951555
Los Angeles, CA 90095-1555}

\date{\today}

\begin{abstract}

Experiments on $^4$He films reveal an attractive Casimir-like force at the bulk $\lambda$-point, and in the superfluid regime.  Previous work has explained the magnitude of this force at the $\lambda$ transition and deep in the superfluid region but not the substantial attractive force immediately below the $\lambda$-point. Utilizing a simple mean-field calculation renormalized by critical fluctuations
we obtain an effective Casimir force that is qualitatively consistent with the scaling function  $\vartheta$ obtained by collapse of experimental data. 

\end{abstract}

 \maketitle

Experiments by Garcia and Chan \cite{chan1999} have challenged our understanding of finite size behavior in thin films of $^4$He close to the $\lambda$-point. 
The experiments monitor the thickness of a wetting film of helium suspended above a $^4$He  bath.
Near and below the bulk transition temperature the film becomes thinner. Since the microscopic intermolecular interactions are not expected to be modified at the  $\lambda$-temperature, the thinning can only be due to collective behavior at and below the transition. Indeed, the reduced thickness has been interpreted as due to an attractive, fluctuation-induced, Casimir-like force. Subsequent results by Ganshin, {\em et. al.} \cite{chan2006} confirm that this force is consistent with finite size scaling, according to which \cite{fisherdegennes,fb}, for a slab of width, $L$,  the effective force per unit area $f(T,L)$, acting to thicken or thin the film, has the general form
\begin{equation}
f(T,L) = \frac{k_BT_c}{L^{d}} \vartheta(tL^{1/ \nu}).
\label{eq:casforce1}
\end{equation}
Here, $t=(T-T_c)/T_c$ is the reduced temperature, measuring the distance from the bulk critical point $T_c$,  $\vartheta$ is the dimensionless amplitude of the scaling function, and $\nu$ is the correlation length exponent. 
\begin{figure}[htbp]
\begin{center}
\includegraphics[width=3in]{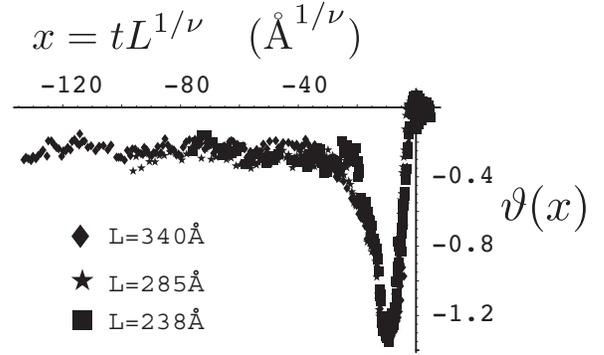}
\caption{The Casimir force amplitude $\vartheta$ for $^4$He films of three different thickness $L$,  as a function of the scaling variable $x=tL^{1/\nu}$, from the data by Ganshin, et. al. \cite{chan2006,garciathanks}. 
Despite the differences in film thickness, the scaled forces collapse onto a single curve with a minimum at $x=-9.7 \pm 0.8 {\rm  \AA^{1/\nu}}$.}
\label{fig:chanplot}
\end{center}
\end{figure}
Figure \ref{fig:chanplot} shows  the experimental data for the dependence of the function
$\vartheta(x)$ on  the scaling combination $x= tL^{1/ \nu}$, as presented in Ref. \cite{chan2006} (see also Fig. 3 in that article). This plot combines the results from $^4$He films of three different initial thicknesses of 238, 285, and 340  ${\rm  \AA}$.  It is important to note that these films are in equilibrium with the bulk Helium liquid.  

As shown in the figure, the force amplitudes extracted from the data associated with the three films collapse onto a single curve, in  clear vindication  of finite size scaling. The behavior of the scaling function {\sl at and immediately above} the $\lambda$ transition is well described by the two-loop renomalization group calculations of Krech and Dietrich\cite{dietrich1,dietrich,krech2}.  Another feature that is successfully explained by theory is the finite negative value of $\vartheta$ well {\sl below the critical point}:
it was demonstrated in Ref.~\cite{zandi} that a combination of phase fluctuations within the film
and surface deformations can account for this value. 

However, the most prominent feature  in Fig.~\ref{fig:chanplot} is the deep minimum of the scaling function   in the vicinity of 
\begin{equation}
x \equiv tL^{1/ \nu}=-9.7 \pm 0.8 \ {\rm  \AA^{1/\nu}} .
\label{eq:xeq}
\end{equation}
\noindent The amplitude of the force at this minimum is roughly an order of magnitude larger than that of the Casimir forces at ($x=0$) and far below ($x\to-\infty$) the transition and has thus far not been reproduced by any theoretical calculation. 
In particular,  the dip is considerably larger than the one in the simulations reported by Dantchev and Krech \cite{dk}, or the low temperature vortex loop calculations of Williams \cite{williams1,williams2,dietrich2}. 

In this letter, we show that the origin of the large attractive Casimir-like force measured in Refs. \cite{chan1999,chan2006} is twofold. First, the boundary conditions that apply in the case of a superfluid film are Dirichlet in that the superfluid order parameter effectively vanishes at the substrate and at the liquid-vapor interface. This distinguishes the wetting film from the periodic systems considered in Refs. \cite{williams1,williams2,dk}. Second, because the film is in equilibrium with a reservoir of bulk fluid, it is necessary to take into account the free energy of the bulk in calculating the effective force, $f$, acting on the film. This leads to the following scenario on cooling the system: Immediately below the $\lambda$ temperature, ordering in the form of a non-zero superfluid order parameter takes place in the bulk reservoir of $^4$He. However, boundary effects suppress the appearance of superfluid ordering, even at a local level, in the film. To take advantage of the lower free energy of the superfluid state, helium atoms move from the film to the ordered bulk, leading to additional film thinning, and a corresponding increase
in the attractive effective Casimir-like force.  Upon further reduction of temperature, local ordering finally occurs in the film; its free energy effectively ``recovers'' to a value similar to the bulk,  resulting in a rebound of the film thickness.

Both of these contributing factors are well-displayed in a simple mean field treatment augmented by the renormalizing effects of critical fluctuations at the onset of order in the bulk. Calculations based on this approach indicate that the minimum of the scaling function occurs when the dimensionless quantity $ {y}=(L/\xi)^{1/ \nu}$,
where $\xi$ the correlation length, is at $y_{\rm min}=-\pi^2 = - 9.87$. 
The experimental value of  $y_{\rm min}^x$ is equal to $-7.4 \pm 0.6$, as is obtained by setting $t=(\xi_0/\xi)^{1/\nu}$ in Eq.~\ref{eq:xeq}, with $\xi_0=1.2 \ {\rm \AA}$ \cite{pobell}.  We note at the outset that the mean-field calculation, details of which follow, 
yields a minimum amplitude which is significantly greater than the observed value. We are optimistic that a proper treatment which correctly accounts for the effects of fluctuations---and for the fact that the superfluid order parameter should be modeled by a two-component vector---will lead to a substantial \emph{reduction} in the force. However, it is clear that any proper theoretical model will have to embody the elements underlying the results reported here. 

The simplest depiction of  the superfluid transition in a thin film geometry is based on the Ginzburg--Pitaevskii Hamiltonian\cite{gp}, which follows from the Ginzburg--Landau model\cite{gl}. In one dimension (moving through the film) the effective Hamiltonian  is
\begin{equation}
\mathcal{H} = \int_0^L \left\{ \frac{1}{2}\left( \frac{d \phi(z)}{dz}\right)^2 + \frac{r}{2}\phi(z)^2 + u \phi(z)^4\right\} dz,
\label{eq:ham1}
\end{equation}
where $\phi(z)$ is the order parameter, $r$ is the reduced temperature, and $u$ is the fourth order coupling constant, which quantifies the system's self interaction and provides the mechanism by which ordering saturates.

The extremum (saddle point or mean field) equation satisfied by the order parameter is
\begin{eqnarray}
 0 = \frac{\delta \mathcal{H}}{\delta \phi(z)} &= & -  \frac{d^2 \phi(z)}{dz^2} +  r \phi(z) + 4 u \phi(z)^3 .\nonumber 
\label{eq:eos1}
\end{eqnarray}
We assume Dirichlet boundary conditions, $\phi(0)=\phi(L)=0$, at both sides of the film consistent with the experimental situation. For the bulk ($L\to\infty$), or in a periodic and hence homogeneous system, a non-zero solution appears as soon as $-r > 0$.  For the film, it is easy to see that a non-zero solution appears only for $-r >\lambda_0$, where $\lambda_0=\pi^2/L^2$ is the smallest (in magnitude) eigenvalue of the Laplacian with Dirichlet boundary conditions.

By quadratures, the full solution to this equation is
\begin{equation}
\frac{1}{2}\left(\frac{d \phi(z)}{dz} \right)^2 = -\frac{r}{2}\left(\phi_0^2 -  \phi(z)^2\right) + u \left( \phi(z)^4 - \phi_0^4\right),
\label{eq:eos2}
\end{equation}
where we have obtained Eq.~(\ref{eq:eos2}) from Eq.~(\ref{eq:ham1}) by exploiting the mathematical equivalence between the first version of the equation of state and Newton's second law for the motion of a particle in a one-dimensional potential. Equation~(\ref{eq:eos2}) is the statement of conservation of energy in this context. 

In accord with the boundary conditions and obvious symmetry considerations, 
we set $\phi^{\prime}(L/2) =0$ and thus $\phi(L/2) = \phi_0$. Integrating $dz/d\phi$ in Eq.~(\ref{eq:eos2}), from the edge of the film to its midpoint gives \cite{dietrich3}
\begin{eqnarray}
\frac{L}{2}& = & \int_0^{\phi_0} \frac{d \phi}{\sqrt{-r(\phi_0^2-\phi^2)[1+(2u/r)(\phi_0^2+\phi^2)]}} \nonumber \\
& = & \frac{1}{\sqrt{-r}}\frac{K\left(\frac{\eta}{1-\eta}\right)}{\sqrt{1-\eta}},
\label{eq:Leq}
\end{eqnarray}
where $K$ is the elliptic integral of the first kind and
\begin{equation}
\eta = \frac{2u \phi_0^2}{-r}.
\label{eq:etaeq}
\end{equation}

To obtain the effective force, we calculate the derivative of the free energy in Eq.~(\ref{eq:ham1}) with respect to $L$, after replacing $\phi(z)$ with its extremum value.
After a series of relatively straightforward steps, we end up with the following result for the $L$-derivative of the free energy
\begin{equation}
\frac{\partial \mathcal{F}}{\partial L} =-\left.\frac{1}{2} \left( \frac{d \phi(z)}{dz} \right)^2 \right|_{z=L} +\frac{r}{2}\phi(L)^2 + u \phi(L)^4 .
\label{eq:Ld3}
\end{equation}
Finally, making use of the Dirichlet boundary conditions, and Eq.~(\ref{eq:eos2}), we have for the derivative of the free energy with respect to the thickness of the system,
\begin{equation}
\frac{\partial \mathcal{F}}{\partial L} = \frac{r}{2}\phi_0^2 + u \phi_0^4 .
\label{eq:Ld5}
\end{equation}
We note  that Eq.~(\ref{eq:Leq}) allows us to write $\eta$ as a function of the combination $\sqrt{-r}L/2$. Setting $r = y/L^2$ and making use of Eq.~(\ref{eq:etaeq}), the derivative of the free energy with respect to $L$ takes the form
\begin{equation}
\frac{\partial \mathcal{F}}{\partial L} =-\frac{1}{4uL^4} y^2 \eta(\sqrt{-y}/2) (1-\eta(\sqrt{-y}/2)).
\label{eq:Ld6}
\end{equation}
Except for the quantity $u$ and the prefactor of $1/L^4$; which is appropriate to a \emph{four} dimensional system (and is consistent with the fact that mean field theory agrees with 4--dimensional hyperscaling); we have a free energy derivative that depends on the scaling variable $y$. 
Within mean field theory, the quantity $u$  is material-dependent.  However, the existence of a stable renormalization group fixed point on the critical hypersurface implies a universal value for this coefficient \cite{wilson}.

\begin{figure}[htbp]
\begin{center}
\includegraphics[width=3in]{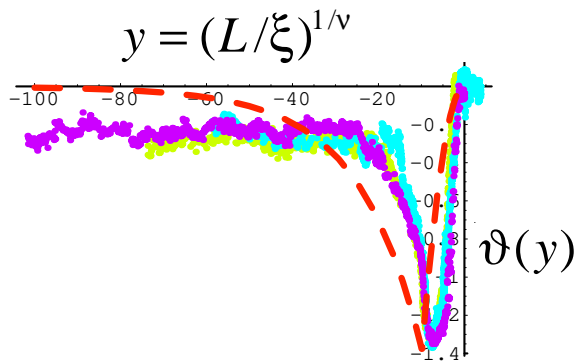}
\caption{The mean field scaling function $\vartheta(y)$ derived from Eq.~(\ref{eq:mfdip1}) (dashed curve), compared to the re-plotted experimental data from Fig.~\ref{fig:chanplot}. 
The horizontal axis is dimensionless and provides a check of the approximations, while
the vertical axis is adjusted so that the minima occur at the same point.}
\label{fig:compdip1}
\end{center}
\end{figure}

An important feature of the experiments in Refs.~\cite{chan1999,chan2006} is that the wetting film is in equilibrium with the vapor above a bulk reservoir of  $^4$He. Thus, to determine the layer thickness we should calculate the change in free energy as some fluid is removed from the film and simultaneously added to the bulk.
The mean field calculation of the bulk free energy is straightforward, obtained from the previous result in the limit of $L \rightarrow \infty$. The resulting expression for the  force is
\begin{equation}
f= 
\frac{1}{4uL^4} y^2 \eta(\sqrt{-y}/2) (1-\eta(\sqrt{-y}/2))-\frac{1}{16uL^4}y^2\ .
\label{eq:mfdip1}
\end{equation}
Figure \ref{fig:compdip1} displays the above mean field scaling function, $\vartheta(y)={fL^4/k_BT_c}$, as a function of $y=rL^{2}=(L/\xi)^2$, where $\xi$ is the correlation length as expressed before. Also shown in the figure is the data in Fig.~\ref{fig:chanplot}. Note that in Fig.~\ref{fig:chanplot} the scaling function is plotted vs. $x$ which has units of ${\rm  \AA^{1/\nu}}$, while in Fig.~\ref{fig:compdip1} the horizontal axis represents the dimensionless quantity $y=(L/\xi)^{1/\nu}$.
The amplitude of the mean field function depends on the parameter $u$, and has been adjusted to give the observed amplitude at the minimum. %The horizontal axis for the experimental data is the scaling combination $y=tL^{1/ \nu}$, but otherwise there is no adjustment of this axis.
%It is thus significant that location of the minimum in the data is close to the $x_{\rm min}=-\pi^2$.
Note that at this coarse resolution, the mean-field curve captures the main trends of the experimental data below the critical temperature. The most significant discrepancy is that the measured force does not go to zero for $y\to-\infty$; this is due to the absence of Goldstone and surface modes in the mean-field analysis\cite{zandi}.

To obtain the correct scaling of the force $f$ with the film thickness $L$, we utilize a version of mean field theory that takes into account the renormalizing effects of critical fluctuations (see Ref.~\cite{rn}). 
The relevant free energy density, after renormalizing to cell blocks larger by
a factor $b=e^{l^*}$, is
\begin{equation}
e^{-l^*d}\left[\frac{r(l^*)}{2} \phi(l^*)^2 + u(l^*) \phi(l^*)^4 \right].
\label{eq:rmf1}
\end{equation}
Close to a fixed point of the renormalization group, the various quantities in the above equation 
scale as~\cite{rn} $r(l^*)  =  re^{l^*/\nu} $, $u(l^*) = u^*$, and $\phi(l^{*})  =  \phi e^{(d-2+ \eta)l^*/2}$. 
To account for the finite thickness of the wetting films, we choose a rescaling factor $e^{l^*}=L$,
at which point the Landau-Ginzburg parameters are $r(l^*) =rL^{1/ \nu}$ and $u(l^*) \approx u^*$.
Ignoring fluctuations at scales larger than (and equal to) $L$ is then equivalent to carrying out
the previous mean-field calculations with the above rescaled parameters (and with $L'=1$).
Any free energy density (and hence effective force) in this approximation is thus obtained from its
mean-field form as
\begin{equation}
f_{\rm ren}(r,u,L)=\frac{1}{L^d}f_{\rm mean-field}\left((L/\xi)^{1/ \nu},u^*,1\right).
\label{eq:rmfx}
\end{equation}
In particular, from Eq.~(\ref{eq:mfdip1}), the effective force for $-\pi^2\leq y\leq 0$ becomes
\begin{equation}
f_{\rm bulk}= - \frac{(L/\xi)^{1/ \nu}}{16u^*L^d}.
\label{eq:rmf3}
\end{equation}

The above mean-field approach with renormalized parameters, also fixes the vertical scale in 
Fig.~\ref{fig:compdip1}, removing the unknown value of $u$. This is because, in units where $k_BT_c=1$, the fourth order coupling constant $u^*$ has a specific value within renormalization group, and is computed to high precision in Ref.~\cite{nickel}. Making use of the values provided there\cite{uvalues}, one finds $u^*=0.88$.  Using Eq.~(\ref{eq:rmf3}), the minimum of the scaling function at $y_{\rm min}=-\pi^2$ is estimated as
\begin{eqnarray}
\vartheta_{\rm min} =-\frac{1}{ .88} \times \frac{(\pi^2)^2}{16} = -6.92,
\label{eq:vthetaval}
\end{eqnarray}
which is roughly five times larger than the experimentally measured value!

The mean-field calculation suffers from a number of other shortcomings.  We already noted that the force is asymptotically zero for $y\to-\infty$ due to the neglect of phase and capillary fluctuations~\cite{zandi}. It also yields a zero critical Casimir force for all $y\geq 0$, while there is a small finite value to this quantity in experiments due to order parameter fluctuations, as calculated at one and two loop level in Ref.~\cite{krech2}. As a first correction to the mean-field result the Casimir force can be estimated by including quadratic (i.e. one-loop) fluctuations around the saddle point. At exactly $y=0$, the fluctuations are mass-less and lead (in three dimensions) to a Casimir
amplitude of~\cite{krech2,inprogress}
\begin{equation}
\vartheta(0)\approx -\frac{1}{2 \pi} \int_0^{\infty}dq \  q^2 \left( \coth q-1\right)   =  -0.0956566.
\label{eq:Gauss}
\end{equation} 
We note that the ratio of the extremum value of $\vartheta(-\pi^2)$ (a mean field result) to its one-loop magnitude at $y=0$ is  around 70. This is to be contrasted with ratios from experimental data that range from 20 to approximately 35 (the variation of possible ratios arises from the different values of $\vartheta(0)$ that can be inferred from experimental data).

Another feature of the mean-field result is the discontinuous slope of the scaling function at its minimum in Fig.~\ref{fig:compdip1}. This is a consequence of the onset, in the mean field approximation, of an
actual transition accompanied by long-range order in the film at $y=-\pi^2$. Calculations involving lower dimensional models for the ``bulk'' and ``film'' (e.g. two dimensional Ising bulk and a one dimensional Ising film or one dimensional bulk and zero dimensional film) yield behavior that is qualitatively similar to the mean field picture notwithstanding that a transition within the film is precluded \cite{inprogress}.  In these cases---as may well be the case in the experimental setups of \cite{chan1999} and \cite{chan2006}---the free energy of the film begins to decrease due to the formation of local pockets of order.   In the absence of a genuine  transition in the film there is no singularity at the minimum.  In the actual films, the onset of superfluidity (which should occur at a different temperature altogether) will be of the two-dimensional $XY$ universality class~\cite{hohenberg,kt}.
These details are certainly not taken into account in our treatment and, arguably, are not relevant for the current level of experimental resolution.

Indeed, let us reemphasize that the essence of our derivation is thermodynamic: outside a narrow ``critical window'' thermodynamic signatures of ordering (whether ultimately short or long--ranged) should be accurately described by a Landau theory.  A phenomenological calculation of the thinning effect may be performed on this basis using standard mean field approximation, The predicted thinning of the film is consistent with the experimental results \cite{inprogress}.

In summary, we find that a mean field calculation yields results for the scaling form of the Casimir force in $^4$He films that are in qualitative agreement with recent experimental observations. While discrepancies remain to be resolved, we are confident that this approach captures the important thermodynamic signatures, and therefore the effective thinning force in $^4$He films at the onset of bulk superfluidity.  Work on an improved mean field approaches and the proper evaluation of the effects of fluctuations in all temperature regimes is ongoing. An understanding the underlying principles behind the thinning of Helium films is expected to have important implications in the analysis of other wetting experiments~\cite{balibar,ueno}.

 The authors would like to acknowledge helpful discussions with Professors R. Garica and M. H. W. Chan, and thank them for the data displayed in Fig.~\ref{fig:chanplot}.  We are additionally grateful to Professor G. Williams and Dr. D. Dantchev for illuminating discussions as well as  Professor T. Tao for some mathematical consultation. Financial support from NSF grant numbers DMR-06-45668 (R.Z), DMR-04-26677 (M.K.), DMS-03-06167 (L.C), and DMR-04-04507 (J.R.) is gratefully acknowledged.

\bibliographystyle{apsrev}
\bibliography{dip}

\begin{thebibliography}{26}
\bibitem[{uva()}]{uvalues}
\bibinfo{note}{See especially the line below Eq. (2.4) in \cite{nickel} and
  Table XII in that article.}
\end{thebibliography}

\end{document}